\documentstyle[12pt]{article}      
\begin{document}

\title{Replica analysis of the $p$-spin interactions \\
 Ising  spin-glass model}

\author{ Viviane M de Oliveira and J F Fontanari \\
Instituto de F\'{\i}sica de S\~ao Carlos\\
Universidade de S\~ao Paulo\\
Caixa Postal 369\\
13560-970 S\~ao Carlos SP\\
Brazil}

\date{}

\maketitle

\centerline{\large{\bf Abstract}}

\bigskip

The thermodynamics 
of the infinite-range Ising spin glass  with $p$-spin 
interactions in the presence of an external magnetic field $h$
is investigated  analytically using the replica method.
We give emphasis to the analysis of the transition between 
the replica symmetric and the one-step replica symmetry 
breaking regimes. In particular, we derive analytical conditions
for the onset of the continuous transition, as well as for 
the location of the
tricritical point at which the transition between those two
regimes becomes discontinuous.

\bigskip

\bigskip

{\bf Short Title:}  $p$-spin Ising model


\newpage


\small

\section{Introduction}\label{sec:level1}

Although the thermodynamics of the Ising spin glass with 
infinite-range interactions, so-called 
Sherrington-Kirkpatrick (SK) model \cite{SK}, has been thoroughly
investigated in the last two decades \cite{Binder,Mezard},
comparatively little attention has been given to the
analysis of
a natural generalization of the SK model, namely,
the $p$-spin interactions Ising spin glass. This model is
described by the Hamiltonian \cite{Derrida,Gross}
\begin{equation}\label{H_p}
{\cal{H}}_p \left ( {\bf S} \right ) = 
- \sum_{1 \leq i_1 <i_2 \ldots < i_p \leq N}
J_{i_1 i_2 \ldots i_p} \, S_{i_1} S_{i_2} \ldots S_{i_p}
- h \, \sum_i S_i
\end{equation}
where $S_i = \pm 1, i=1,\ldots,N$ are Ising spins and $h$ is the  
external magnetic field. Here
the coupling strengths are statistically independent 
random variables with a Gaussian distribution
\begin{equation}\label{prob}
{\cal{P}} \left ( J_{i_1 i_2 \ldots i_p} \right ) =
\sqrt{\frac{N^{p-1}}{\pi p!}} \exp \left [ 
-\frac{ \left( J_{i_1 i_2 \ldots i_p} \right)^2 N^{p-1}}{p!} 
\right ] .
\end{equation}
Besides the acknowledged importance of the
$p$-spin interactions Ising spin glass in the framework of
the traditional statistical mechanics of disordered systems 
(it yields the celebrated random 
energy model in the limit $p \rightarrow \infty$ \cite{Derrida}
and the SK model for $p=2$), it also plays a significant role in the
study of adaptive walks in rugged fitness landscapes
within the research program championed by Kauffman 
\cite{Kauffman,Amitrano,Weinberger}.

The thermodynamics of the SK model ($p=2$) as well as that
of the random energy model ($p \rightarrow \infty$) are now
well understood. 
In particular, for $p=2$ and $h=0$ the order parameter 
function $q(x)$ tends to
zero continuously as the temperature approaches  the critical
value $T_c^{(2)} = 1$ at which the  transition  between 
the spin-glass and the high temperature (disordered)  phases 
takes place \cite{Binder,Mezard}. 
For $p \rightarrow \infty$ and $h=0$, the system has 
a critical temperature 
$T_c^{(\infty)} = 1/\left ( 2 \sqrt{\ln 2} \right )$ 
at which it freezes  completely into the ground state: 
$q(x)$ is a step function with values zero
and one, and with a break point at 
$x = T/T_c^{(\infty)}$ \cite{Derrida,Gross}. 
These results are not affected qualitatively by the presence
of a non-zero magnetic field. In particular, for
the SK model the critical temperature decreases
monotonically with increasing $h$ while the transition remains
continuous, in the sense that $q(x)$ is continuous at 
the transition line \cite{AT,Parisi}. In contrast, 
for the random energy model
the critical temperature increases with increasing  $h$ while 
the discontinuity in the step function $q(x)$ decreases with 
increasing $h$ and vanishes in the limit $h \rightarrow \infty$
\cite{Derrida,Gross}.

The situation for finite $p > 2$ is considerably more complicated
and so the thermodynamics of the $p$-spin model has been
investigated  for $h=0$ only 
\cite{Gardner,Stariolo}. In this case
there is a transition from the disordered phase to a partially
frozen phase characterized by a step function $q(x)$ with values
zero and $q_1 < 1$. As the temperature is lowered further, a
second transition occurs, leading to a phase described by
a continuous order parameter function \cite{Gardner,Stariolo}. 
However, there are evidences that the presence of a
non-zero magnetic field decreases the size of the discontinuity
of the order parameter $q(x)$ leading, eventually, to a 
continuous phase transition.  In fact, a recent analysis of the typical
overlap $\bar{q}$ between pairs of metastable states with energy density 
$\epsilon$ indicates that  $\bar{q}$ is a discontinuous function 
of $\epsilon$ for $p > 2$, and that the size of the jump in $\bar{q}$ 
increases with $p$ and decreases with  $h$,
vanishing at finite values of the magnetic field \cite{Viviane}.
Moreover, a similar effect has already
been observed in the thermodynamics analysis of the 
spherical $p$-spin interaction spin-glass model 
\cite{Cris}. It is interesting to note that the spin-glass phase 
of this continuous spin model is described exactly by a step order 
parameter function, i.e., the one-step replica symmetry breaking
is the most general solution within the Parisi scheme of replica 
symmetry breaking \cite{Cris}.

In this paper we use the replica method to study the 
thermodynamics of the Ising $p$-spin interaction spin-glass 
model in the presence of the magnetic field $h$. 
We focus on the effects of $h$ on the transition between the
replica symmetric (RS) and 
the one-step replica symmetry breaking (1 RSB) regimes. 
In particular, we show that for $p > 2$ the
discontinuous transition reported in previous analyses
\cite{Gardner,Stariolo} turns into a continuous one for $h$
larger than a certain value  $h_T$. Moreover, we 
derive analytical conditions to determine the location
of the continuous transition line, as well as that of
the tricritical point at which the transition becomes discontinuous.

The remainder of the paper is organized as follows. In
Sec.\ \ref{sec:level2} we discuss  the replica formulation
and present the formal equation for the average free-energy density,
which is then rewritten using the RS and the 1 RSB 
{\it ans\"{a}tze}. These results are discussed very briefly
since their derivations are given in detail in  Gardner's paper
\cite{Gardner}. We also present the solution of the 1 RSB 
saddle-point equations  in the limit of large $p$,
extending thus the series expansions results  for non-zero $h$.
In Sec.\ \ref{sec:level3} we derive analytical conditions
for locating the continuous transition and the
tricritical point between the RS and 1 RSB regimes, and
present the phase
diagrams in the plane $(T,h)$. Finally, some concluding remarks 
are presented in Sec.\ \ref{sec:level4}.

%
%
\section{The replica formulation}\label{sec:level2}

We are interested in the evaluation of the average free-energy
density $f$ defined by
\begin{equation}\label{f_1}
- \beta f = \lim_{N \rightarrow \infty} \frac{1}{N}~ 
\langle \langle \ln Z \rangle \rangle
\end{equation}
where
\begin{equation}\label{Z}
Z = \mbox{Tr}_{\bf S} 
\exp \left [ -\beta {\cal{H}}_p \left ( {\bf S} \right ) \right ],
\end{equation}
is the partition function and $\beta$ is the inverse temperature.
Here $\langle \langle \ldots \rangle \rangle$ stands for the 
average over the
coupling strengths, and  $\mbox{Tr}_{\bf S}$ denotes the summation 
over the $2^N$  states of the system.
As usual, the evaluation of the quenched
average in Eq.\ (\ref{f_1}) can be effectuated through the
replica method: using the identity
\begin{equation}
\langle \langle \ln Z \rangle \rangle = 
\lim_{n \rightarrow 0} \frac{1}{n}~
\ln ~\langle \langle Z^n \rangle \rangle
\end{equation}
we first calculate $\langle \langle Z^n \rangle \rangle$ 
for {\em integer}
$n$, i.e. $Z^n = \prod_{a=1}^n Z_a$,
and then analytically continue to $n=0$ \cite{Binder,Mezard}.
The final result is simply \cite{Gardner}
\begin{equation}\label{f_2}
- \beta f = \lim_{n \rightarrow 0}  \mbox{extr} ~ \left [ \frac{1}{n} 
G \left( q_{ab}, \lambda_{ab} \right ) \right ] + \frac{1}{4}\beta^2
\end{equation}
where
\begin{eqnarray}\label{G_ab}
G \left( q_{ab}, \lambda_{ab} \right ) & = &
 \frac{1}{2}\beta^2 \sum_{a < b} q_{ab}^{~p}
- \beta^2 \sum_{a < b} \lambda_{ab} q_{ab} \nonumber \\
& & \mbox{} + \ln \mbox{Tr}_{\{S^a\}} \exp \left ( 
\beta^2 \sum_{a < b} 
\lambda_{ab} S^a S^b + \beta h \sum_a S^a \right ) .
\end{eqnarray}
The extremum in Eq.\ (\ref{f_2}) is taken over
the physical order parameter
\begin{equation}
q_{ab} = \langle \langle \frac{1}{N} 
\sum_{i=1}^N \langle  S_i^a \rangle_T  \langle S_i^b \rangle_T
\rangle \rangle  ~~~~~ a < b,
\end{equation}
which measures the overlap between two different equilibrium states
${\bf S}^a$ and ${\bf S}^b$, and over its corresponding 
Lagrange multiplier $\lambda_{ab}$. Here, 
$\langle \ldots \rangle_T$ stands for a thermal average.
To proceed further, next we consider two standard {\it ans\"{a}tze}
for the structure of the saddle-point parameters.

\subsection{Replica symmetric  solution}
In this case we assume that the saddle-point parameters are
symmetric under permutations of the replica indices, i.e.,
$q_{ab} = q$ and $\lambda_{ab} = \lambda$. With this prescription
the evaluation of Eq.\ (\ref{f_2}) is straightforward, resulting
in the replica-symmetric free-energy density
\begin{equation}\label{f_rs}
-\beta f_{rs} = -\frac{1}{2} \beta^2 \lambda \left ( 1 - q \right )
+ \frac{1}{4} \beta^2 \left ( 1 - q^p \right ) +
\int_{-\infty}^\infty Dz \ln 2 
\cosh \left [ \beta \Xi_s \left( z \right ) \right ]
\end{equation}
where 
\begin{equation}\label{Xi_s}
\Xi_s = z \sqrt{\lambda} + h
\end{equation}
and
\begin{equation}
Dz = \frac{dz}{\sqrt{2\pi}} ~ \mbox{e}^{-z^2/2}
\end{equation}
is the Gaussian measure. The saddle-point equations 
$\partial f_{rs}/\partial q = 0$ and
$\partial f_{rs}/\partial \lambda = 0$  yield
\begin{equation}\label{l_rs}
	\lambda = \frac{p}{2} q^{p-1}
\end{equation}
and
\begin{equation}\label{q_rs}
	q = \int_{-\infty}^\infty Dz 
 \tanh^2 \left [ \beta \Xi_s \left ( z \right ) \right ],
\end{equation}
respectively. The replica-symmetric solution is locally stable
wherever the Almeida-Thouless condition \cite{AT}, which in
this case is given by
\cite{Gardner}
\begin{equation}\label{stab}
1 - \beta^2  \left ( p - 1 \right )  ~\frac{\lambda}{q} 
\int_{-\infty}^\infty Dz ~
 \mbox{sech}^4 \left [ \beta \Xi_s \left ( z  \right ) 
\right ] > 0,
\end{equation}
is satisfied. In fact, since Eq.\ (\ref{q_rs}) has 
either one or three positive solutions, this stability condition is 
very useful to single out the physical one. In particular, for
$h=0$ the only stable solution is $q=0$.

\subsection{Replica symmetry broken solution}
Following Parisi's scheme \cite{Mezard}, we carry out the first
step of replica symmetry breaking by dividing the $n$ replicas
into $n/m$ groups of $m$ replicas and setting $q_{ab} = q_1$,
$\lambda_{ab} = \lambda_1$ if $a$ and $b$ belong to the same group
and $q_{ab} = q_0$, $\lambda_{ab} = \lambda_0$ otherwise.
The physical meaning of the saddle-point parameters is the following
\begin{equation}
q_{0} = \langle \langle \frac{1}{N} 
\sum_{i=1}^N \langle  S_i^a \rangle_T  \langle S_i^b \rangle_T
\rangle \rangle  ~~~~~ a < b,
\end{equation}
\begin{equation}
q_{1} = \langle \langle \frac{1}{N} 
\sum_{i=1}^N \langle  S_i^a \rangle_T^2 
\rangle \rangle  ~~~~~ ,
\end{equation}
and $m = 1 - \sum_a P_a^2$. Hence $q_0$ is the overlap between
a pair of different equilibrium states, $q_1$ is the overlap
of an equilibrium state with itself ($q_1 \geq q_0$), 
and $m$ is the probability of finding
two copies of the system in two different states ($P_a$ is just
the Gibbs probability measure for the state ${\bf S}^a$).
We note that in the limit $n \rightarrow 0$, the parameter $m$ 
is constrained
to the range $0 \leq m \leq 1$. Using this prescription,
Eq.\ (\ref{f_2}) becomes
\begin{eqnarray}\label{f_rsb}
-\beta f_{rsb} & = & -\frac{1}{4} \beta^2 
\left [ 2 \lambda_1 \left ( 1 - q_1 +
m q_1 \right ) - 2 m q_0 \lambda_0  - 1 
+ \left ( 1 - m \right ) q_1^p
+ m q_0^p   \right ]  \nonumber \\
& & \mbox{} + \ln 2 + \frac{1}{m}
\int_{-\infty}^\infty Dz_0 \ln \int_{-\infty}^\infty Dz_1
 \cosh^m \beta \Xi  
\end{eqnarray}
where 
\begin{equation}
 \Xi = z_1 \sqrt{\lambda_1 - \lambda_0} + 
z_0 \sqrt{\lambda_0} + h .
\end{equation}
The saddle-point equations
$\partial f_{rsb}/\partial q_k = 0 $ yield
\begin{equation}\label{lambda_k}
\lambda_k = \frac{p}{2} q_k^{p-1} 
\end{equation}
for $k =0,1$. The saddle-point parameters
$q_0$, $q_1$ and $m$ are given by the equations
\begin{equation}\label{q_0}
q_0 = \int_{-\infty}^\infty Dz_0 ~ 
\left \langle \tanh \beta \Xi \right \rangle_z^2 ,
\end{equation}
\begin{equation}\label{q_1}
q_1 = \int_{-\infty}^\infty Dz_0 ~ 
\left \langle \tanh^2 \beta \Xi \right \rangle_z
\end{equation}
and
\begin{eqnarray}\label{m}
\frac{1}{4} \beta^2 (p-1) \left ( q_1^p - q_0^p \right ) & = &
- \frac{1}{m^2}~ \int_{-\infty}^\infty Dz_0 ~ 
\ln \int_{-\infty}^\infty Dz_1
 \cosh^m \beta \Xi  \nonumber \\
& & \mbox{} +  \frac{1}{m}
\int_{-\infty}^\infty Dz_0 ~ \left \langle \ln \cosh \beta \Xi 
\right \rangle_z ,
\end{eqnarray}
where we have introduced the notation
\begin{equation}\label{notation}
\left \langle \ldots \right \rangle_z =
\frac{ \int_{-\infty}^\infty Dz_1  \left ( \ldots \right )
\cosh^m \beta \Xi  }
{\int_{-\infty}^\infty Dz_1
 \cosh^m \beta \Xi  }  .
\end{equation}
It is clear from these equations that the replica-symmetric
saddle-point $q_0 = q_1 = q$ is a solution for any value of $m$.
In general, however, the 1 RSB equations will admit a different
solution. In particular, in the limit $p \rightarrow \infty$ 
the solution is $q_0 = \tanh^2 \left ( \beta m h \right )$,
$q_1 = 1$ and $m = \beta_c /\beta$ where 
$\beta_c = 1/T_c^{(\infty)}$ is the solution of the equation
\cite{Gross}
\begin{equation}\label{Tcin}
\mbox{\small $\frac{1}{4}$} \beta_c^2 =   \ln 2 \cosh \beta_c h
- \beta_c h \tanh \beta_c h  .
\end{equation}
Below $T_c^{(\infty)}$, the entropy vanishes  and $m$ sticks to
its maximum value, namely $m =1$, signaling the existence of a frozen
phase in accordance with the physical meaning of $m$ mentioned before.
It is instructive to consider the finite $p$ corrections to 
the infinite-$p$ solution by expanding the 1 RSB equations around 
that solution. Thus, extending the results of Gardner \cite{Gardner}
for non-zero $h$, we find
\begin{equation}\label{q_0_large}
q_0 = \tanh^2 \left ( \beta m h \right ) \left [ 1 + 
 2 \xi_m ~ \mbox{sech} \left ( \beta m h \right ) ~
\frac{\mbox{e}^{-\beta^2 m^2 p/4}}
{\sqrt{\mbox{\small $\frac{1}{2}$} p \beta^2}}
 \right ] ,
\end{equation}
\begin{equation}\label{q_1_large}
q_1 = 1  -  
 \frac{m \xi_m}{1 - m} ~ \mbox{sech} \left ( \beta m h \right )~
\frac{\mbox{e}^{-\beta^2 m^2 p/4}}
{\sqrt{\mbox{\small $\frac{1}{2}$} p \beta^2}}
\end{equation}
and 
\begin{equation}\label{m_large}
\mbox{\small $\frac{1}{4}$} \beta^2  =    
\frac{1}{m^2} \left [ \ln 2 \cosh \left ( \beta m h \right )
- \beta m h \tanh \left ( \beta m h \right ) \right ] + \Lambda_m 
\end{equation}
where
\begin{equation}
\Lambda_m = - \sqrt{\mbox{\small $\frac{1}{2}$} p \beta^2}~
\xi_m ~ \mbox{sech} \left ( \beta m h \right ) ~
\mbox{e}^{-\beta^2 m^2 p/4}
\end{equation} 
if $ \beta^2 m^2 < 8 \mid \ln \tanh \left ( \beta m h \right ) \mid$
and
\begin{equation}
\Lambda_m = \beta^2 p \tanh^{2p} \left ( \beta m h \right ) 
~\frac{\beta m h}{\sinh \left ( 2 \beta m h \right ) },
\end{equation} 
otherwise. Here,
\begin{eqnarray}
\xi_m & = &  - \frac{1}{\sqrt{2\pi}} \sum_{i=0}^\infty~
\left ( \! \! \begin{array}{c} m \\ i \end{array} \! \! \right )
 ~\frac{1}{2i-m} \nonumber \\
& = & \frac{1}{\sqrt{2\pi}} \int_{-\infty}^\infty dz 
\left [ 2 \cosh \left ( mz \right ) - 2^m \cosh^m \left ( z \right )
\right ] .
\end{eqnarray}
where we have used the extended definition of the binomial coefficient
to real $m$ \cite{Feller}.

At this stage we already can realize the existence of
two solutions of a quite different nature, signaling then
the non-trivial role played by the magnetic field in 
the thermodynamics of the $p$-spin model. 

A quite interesting property of the 1 RSB solution,
which can easily be verified numerically, is that
$q_0 = 0$ for $h=0$ and $p>2$, indicating thus that the
equilibrium states are completely uncorrelated. Moreover, 
this result
has greatly facilitated both the numerical and analytical
analyses of the model, since the integrals over $z_0$ 
in Eqs.\ (\ref{q_0})-(\ref{m}) can be can
carried out trivially in that case \cite{Gardner,Stariolo}.
However, as explicitly shown by
Eq.\ (\ref{q_0_large})  the non-zero magnetic field induces
correlations between different equilibrium states so
that $q_0$ is no longer zero in this case.

For $p=3$, we present in Figs.\ 1, 2 and 3 
the temperature dependence of the  RS and 1 RSB saddle-point 
parameters for $h=0$, $0.5$ and $1$, respectively.
As mentioned before, for $h=0$ we find $q = q_0 = 0$. The size of
the jump in $q_1$  decreases with increasing $h$ and
disappears altogether for $h \geq h_T^{(3)} \approx 0.57$. Of particular 
interest is the temperature dependence of the saddle-point parameter $m$: 
at the discontinuous transition it reaches its maximal value, namely,
$m=1$, while at the continuous transition it assumes a certain value
$m = m_c \leq 1$, which depends on $T$ and $p$. As expected, the behavior 
pattern depicted in Fig.\ 3 is very similar to that found in the
analysis of the magnetic properties of the SK model \cite{Parisi}, as
the transition is continuous in that model.
We note that
since $m$ plays no role in the RS solution, the curve for $m$ must end
at the transition lines.

The location of the transition lines as well as the characterization
of the critical values of the saddle-point parameters are discussed 
in detail in the next section.

\section{Transition lines}\label{sec:level3}

As indicated in the figures presented before, there are two qualitatively
different types of transition between the RS and the 1 RSB regimes
which we will discuss separately in the sequel. 

\subsection{Continuous transition line}
The location of the continuous transition between
the  RS and the 1 RSB solution is determined by solving the 
1 RSB equations in the limit of small $q_1 - q_0$. More
pointedly,
subtracting Eq.\ (\ref{q_0}) from Eq.\ (\ref{q_1}) and
keeping  terms up to the order $\left ( q_1 - q_0 \right )^ 2$
yields 
\begin{equation}\label{q_1-q_0}
q_1 - q_0 = \frac{2 q_0^2}{\beta^2  \left ( p - 1 \right) \lambda_0}
~\frac{B_0 \left ( q_0 \right )}{B_2 \left ( q_0,m \right )}
\end{equation}
where
\begin{equation}
B_0 \left ( q_0 \right ) = 1 - \beta^2  \left ( p - 1 \right ) 
~\frac{\lambda_0}{q_0} 
\int_{-\infty}^\infty Dz ~
 \mbox{sech}^4 \left [ \beta \Xi_0 \left ( z \right ) \right ] ,
\end{equation}
and
\begin{eqnarray}
B_2 \left ( q_0,m \right ) & = &
 \left [  p - 2 + 4 \beta^2 \left ( p -1 \right ) \lambda_0  
\left ( 3 - 2 m \right ) \right ]
\int_{-\infty}^\infty Dz ~
 \mbox{sech}^4  \left [ \beta \Xi_0 \left ( z \right ) \right ]
\nonumber \\
& & - 2 \beta^2  \left ( p -1 \right ) \lambda_0 \left ( 8 - 5 m \right )
 ~\int_{-\infty}^\infty Dz ~
 \mbox{sech}^6 \left [ \beta \Xi_0 \left ( z \right ) \right ] .
\end{eqnarray}
Here
\begin{equation}
\Xi_0 \left ( z \right ) =  z \sqrt{\lambda_0} + h
\end{equation}
with $\lambda_0$ given by Eq.\ (\ref{lambda_k}).
We note that both $B_0$ and $B_2$ are negative quantities in
the 1 RSB regime.
Since at the continuous transition $q_1 \rightarrow q_0 \rightarrow q$, 
where $q$  is the replica symmetric saddle-point (\ref{q_rs}), 
the transition line is given by the condition
\begin{equation}\label{critical}
B_0 \left ( q \right ) = 0
\end{equation}
which, as expected, coincides with the replica-symmetric stability
line given by Eq. ({\ref{stab}). To specify the value of $m$ at
the critical line, denoted by $m_c$, we expand Eq.\ (\ref{m}) 
for small $q_1 - q_0$ (in this case we must keep terms up to the order 
$\left ( q_1 - q_0 \right )^3$) and then subtract it
from Eq.\ (\ref{q_0}).  
Using the condition (\ref{critical}) 
together with $q_0 \rightarrow q$ yields
\begin{equation}\label{m_c}
1 - m_c = - \frac{B_2 \left ( q,1 \right )}
{B_4 \left ( q \right )}
\end{equation}
where
\begin{eqnarray}
B_4 \left ( q \right ) & = &
 12 \left (  p - 2 \right ) ~ \int_{-\infty}^\infty Dz ~
 \mbox{sech}^2  \left [ \beta \Xi_s \left ( z \right ) \right ]
\tanh^2 \left [ \beta \Xi_s \left ( z \right ) \right ]
\nonumber \\
& & + 2 \beta^2  \left ( p -1 \right ) \lambda 
 ~\int_{-\infty}^\infty Dz ~
 \mbox{sech}^6 \left [ \beta \Xi_s \left ( z \right ) \right ] 
\end{eqnarray}
with $\Xi_s$ and $\lambda$ given by Eqs.\ (\ref{Xi_s}) and 
(\ref{l_rs}),
respectively. Eq. (\ref{m_c}) holds 
provided that $m_c  \leq 1$ and so  the continuous transition line must
end at a tricritical point, whose location is obtained by solving
\begin{equation}
B_2 \left ( q,1 \right ) = 0 
\end{equation}
and Eq.\ (\ref{critical}) simultaneously. As usual, 
the denominator in Eq. (\ref{q_1-q_0}) vanishes at the
tricritical point.

\subsection{Discontinuous transition line}

The location of the discontinuous transition line
is determined by equating the  free energies of the RS and
1 RSB solutions, given by Eqs.\ (\ref{f_rs}) and   (\ref{f_rsb}),
respectively. This task is greatly facilitated in this case by
noting that setting $m=1$ in Eq.\ (\ref{q_0}) 
yields $q_0 = q$. Moreover, since for $m=1$ Eq.\ (\ref{f_rsb}) becomes
independent of $q_1$ (and $\lambda_1$) one has $f_{rsb} (m=1) = f_{rs}$.
Thus, for fixed $h$ the temperature at which the discontinuous transition 
takes place is obtained by solving the 1 RSB 
saddle-point equations with $m=1$ for $q_1$, $q_0=q$, and $T=T_c$.

\subsection{Analysis of the results}

The phase diagrams in the plane $(T,h)$ are presented in Figs.\ 4,
5, and 6 for $p=2$, $3$ and $10$, respectively. The solid curves are
the RS stability condition, Eq.\  (\ref{stab}), whose upper 
branch coincides
with the continuous transition line, Eq.\ (\ref{critical}). 
The discontinuous transition lines (dot-dashed curves) 
join the continuous ones at the tricritical points (full circles). 
We also present  the lines
at which the entropy of the RS solution vanishes (short-dashed curves), 
which for $h=0$ intersect the temperature 
axis at $T = 1/\left (2 \sqrt{\ln 2} \right ) \approx 0.60$, 
whatever the value of $p > 2$. We note that for $p \rightarrow \infty$
the condition for the vanishing of the RS entropy yields exactly
the discontinuous transition line for the random energy model, 
Eq.\ (\ref{Tcin}).
The agreement between these lines is already very good for $p =10$
and $h$ not too near $h_t$, as illustrated in Fig.\ 6.

Since our results 
are valid for real, though physically meaningless,
values of $p \geq 2$ as well, we present in Fig.\ 7
the value of the saddle-point parameter $m$ at the continuous
transition line, given by Eq. (\ref{m_c}),  
for several values of $p$. As expected, for $p > 2$ we find 
$m_c = 1$ at the  tricritical points. 
We mention that, despite
the numerous studies of the SK model, we are not aware of any
calculation of the 1 RSB saddle-point parameter $m$ over
the Almeida-Thouless stability line.
In  Figs. 8 and 9  we present
the values of $T$ and $h$ at the tricritical point, respectively, 
as functions of the real variable $p$. For $p \rightarrow 2$ we 
find $h_t \rightarrow 0$ and $T_t \rightarrow 1$, while for large 
$p$ we find that 
$h_t$ increases like $\sqrt{p \ln p}$,
$T_t$ like $\sqrt{p/\ln p}$ and $1 - q_t$ goes to zero like 
$1/\left (p \sqrt{\ln p}\right )$. These results indicate 
that the phase diagrams in the plane $(T,h)$ do not display the 
two types of transitions 
only in the extreme cases $p=2$ and $p \rightarrow \infty$.

\section{Conclusion}\label{sec:level4}

Some comments regarding the validity of
the 1 RSB solution are in order. The stability analysis of
that solution carried out for $h=0$ indicates that it becomes
unstable for low temperatures \cite{Gardner}. A seemly simpler
approach to check the physical soundness of the 1 RSB solution
is to evaluate numerically its entropy.  This procedure, however,
has proved very elusive: since the entropy becomes negative
when it is of order $\mbox{e}^{-\beta p}$, the numerical precision
required to determine the temperature at which it vanishes
$T''$ is
exceedingly large. For instance, for $h=0$ we find 
$T''= 0.10$, $0.087 ~(0.19)$ and
$0.034 ~(0.18)$ for $p=2$, $3$ and $5$, respectively.
The numbers between parentheses are the numerical estimates of
ref.\ \cite{Stariolo}. As our numerical results  are in good
agreement with that of ref.\ \cite{Parisi_2} for $p=2$, and also
are consistent with the trend of decreasing $T''$ 
with increasing $p$, we think they are the correct ones.
Already for $p > 5$, however, we have failed to obtain reliable 
estimates for $T''$. Since the precision problem becomes 
much worse  for non-zero $h$,  due to
the numerical evaluation of the double integrals, 
we refrain from presenting the estimates for $T''$ in that case.

Although for finite $p$ the 1 RSB solution certainly does not  
describe correctly  the low temperature phase of the $p$-spin
Ising spin glass, it probably yields the correct solution near
the transition line delimiting the replica symmetric and the
replica symmetry breaking regimes. In fact, according to Gardner
\cite{Gardner}, considering further steps of replica symmetry 
breaking within Parisi's scheme will result in
a new {\it continuous}  transition between the 1 RSB regime and 
a more complex regime, described by a continuous order parameter
function.
In this sense, we think that our results regarding the 
transition lines between the  RS and the 1 RSB regimes 
are not mere artifacts of the replica method but indeed
describe  genuine features of the thermodynamics of the
infinite range $p$-spin Ising spin glass in a magnetic field.

%

\bigskip

{\bf Acknowledgments}
This work was  supported in part by Conselho Nacional de
Desenvolvimento Cient\'{\i}fico e Tecnol\'ogico (CNPq). 
VMO holds a {\small{FAPESP}} fellowship.




\newpage

\section*{Figure captions}
\bigskip

\parindent=0pt

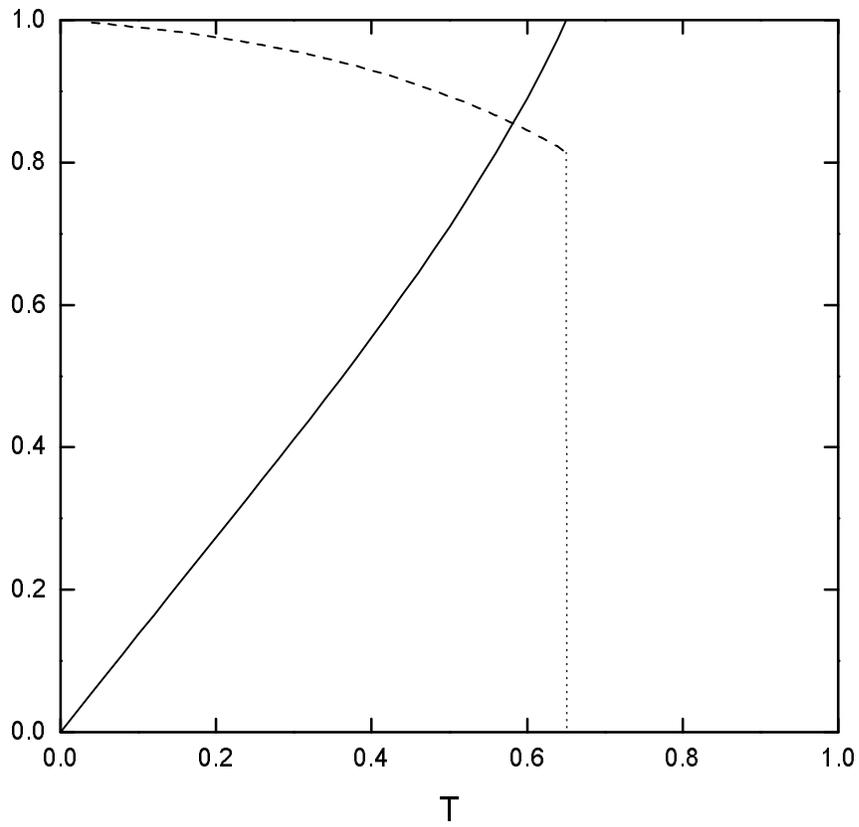
{\bf Fig. 1} One-step replica symmetry breaking saddle-point 
parameters $m$ (solid curve) and $q_1$ (short-dashed curve)
as a function of the temperature $T$ for $p=3$ and $h=0$.
In this case $q_0 = q = 0$. The discontinuous transition occurs at
$T \approx 0.65$.

\bigskip

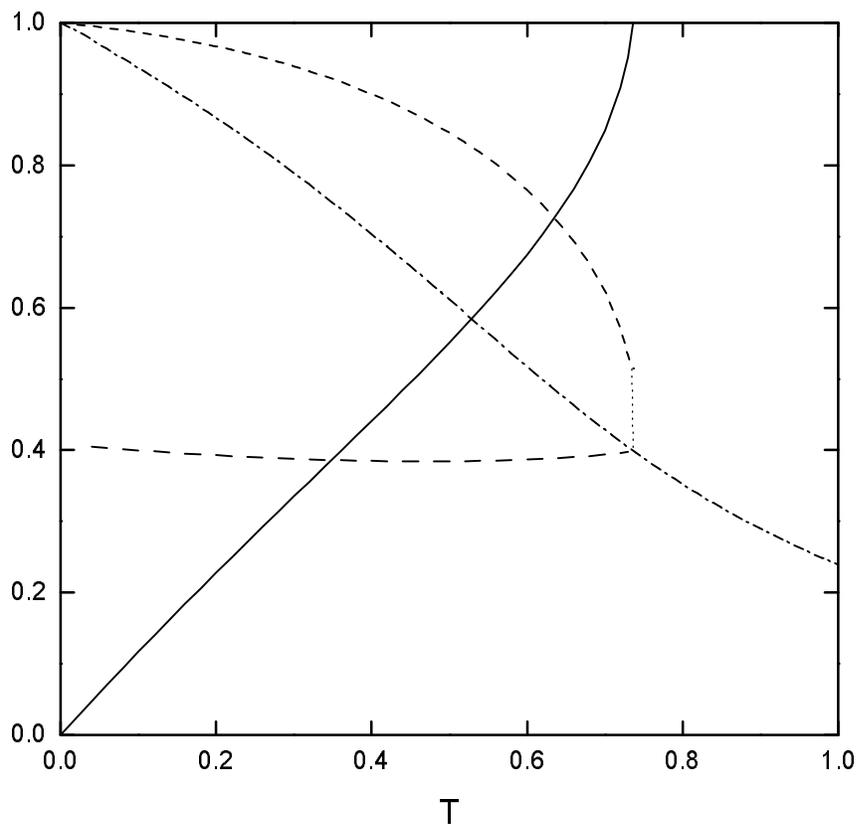
{\bf Fig. 2} One-step replica symmetry breaking saddle-point 
parameters $m$ (solid curve), $q_0$ (long-dashed curve),
$q_1$ (short-dashed curve)
as a function of the temperature $T$ for $p=3$ and $h=0.5$.
The dash-dotted curve is the replica symmetric saddle-point
parameter $q$. The discontinuous transition occurs at
$T \approx 0.74$.

\bigskip

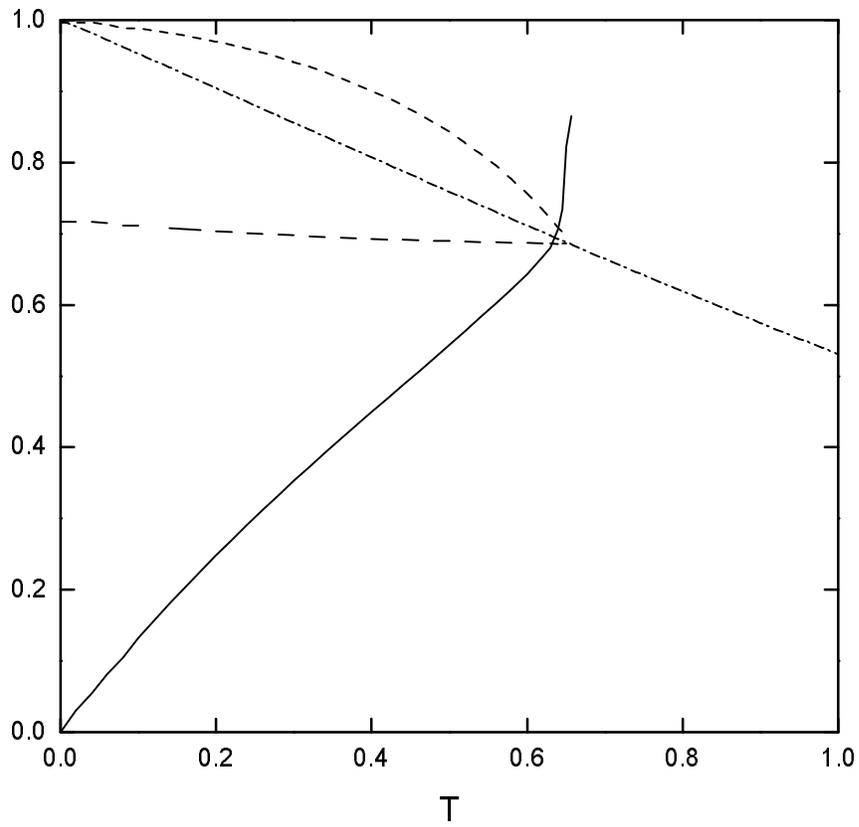
{\bf Fig. 3} Same as Fig.\ 2 but for $h=1$.
The continuous transition occurs at $T \approx 0.66$, at which
$m \approx 0.87$.

\bigskip

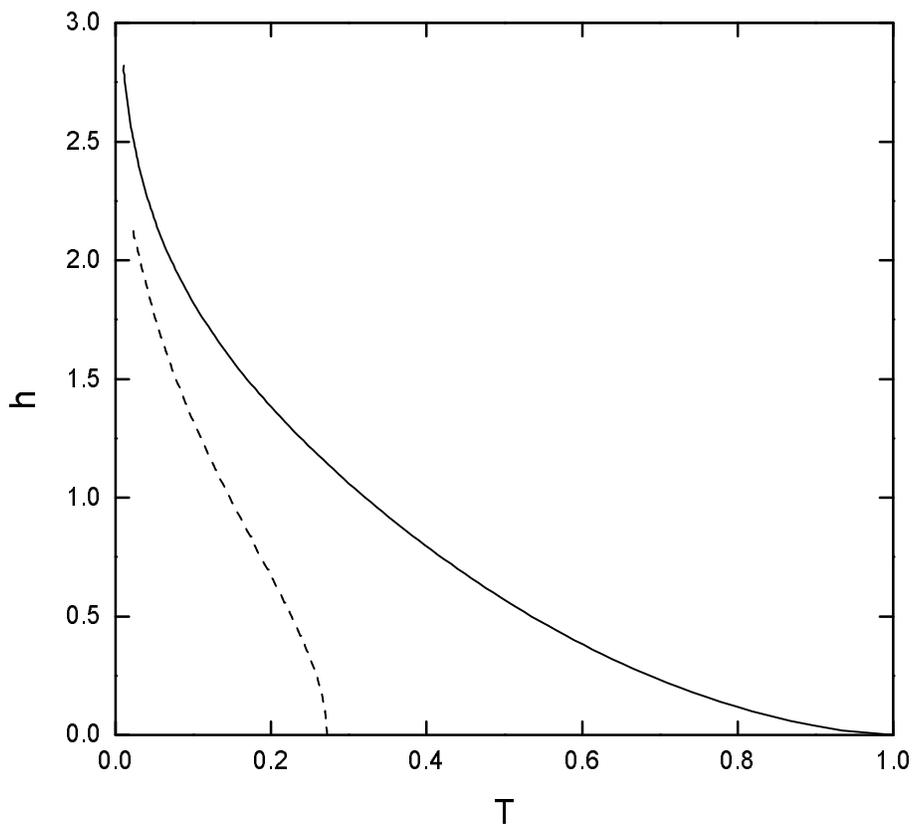
{\bf Fig. 4} Phase diagram in the plane $(T,h)$ for $p=2$. 
The RS saddle-point is locally unstable inside the region
delimited by the solid curve, which coincides with the continuous 
transition line between the RS and the 1 RSB regimes.  The
short-dashed curve delimits the region inside which the
RS entropy is negative.

\bigskip

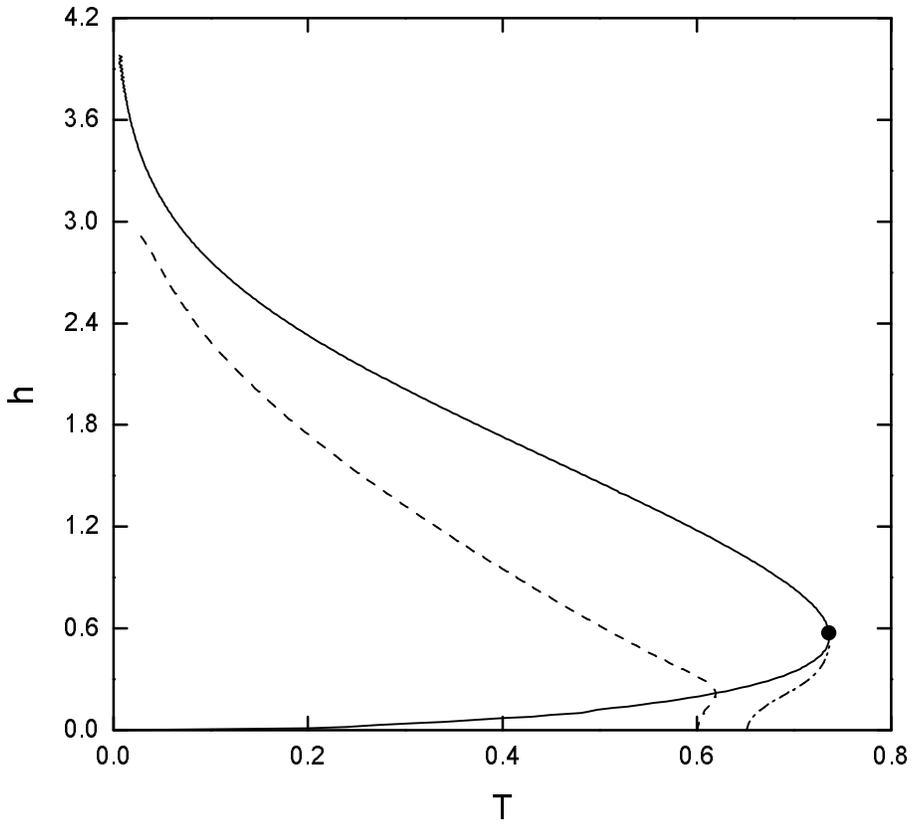
{\bf Fig. 5} Same as Fig.\ 4 but for $p=3$.
The RS stability line coincides with the continuous transition
line in the branch above the tricritical point (full circle),
located at $T_t \approx 0.74$ and $h_t \approx 0.57$.  The
dot-dashed curve is the discontinuous transition line.
The convention is the same as for Fig.\ 4.

\bigskip

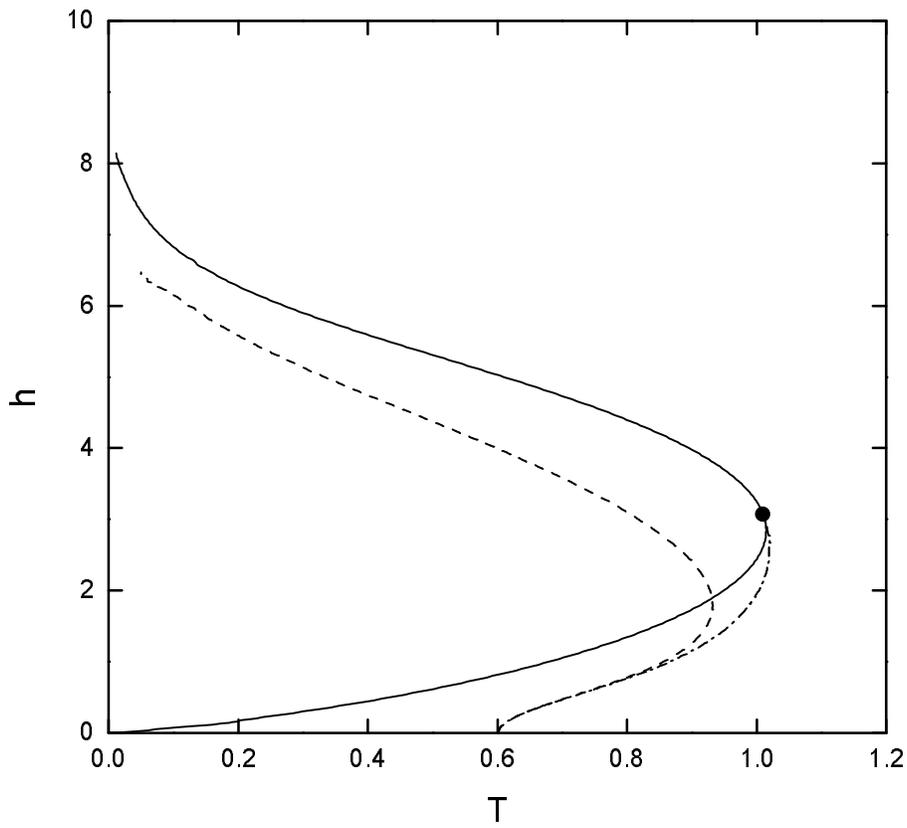
{\bf Fig. 6} Same as Fig.\ 5 but for $p=10$.
The tricritical point (full circle) is
located at $T_t \approx 1.01$ and $h_t \approx 3.07$.  

\bigskip

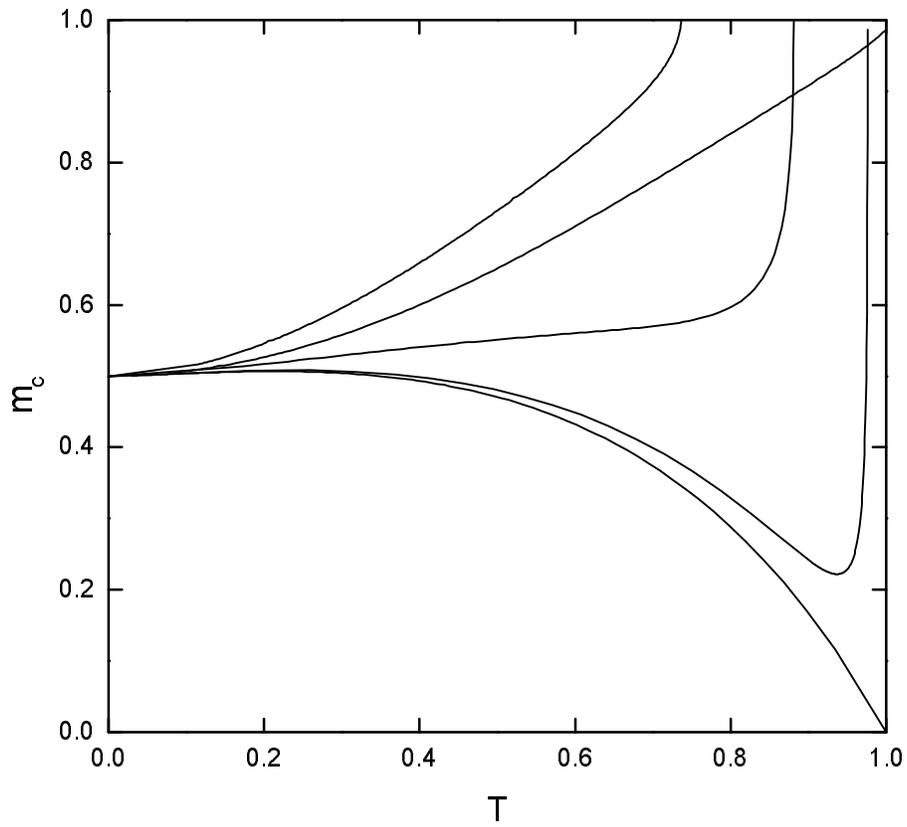
{\bf Fig. 7} Saddle-point parameter $m$ at the continuous
transition line for (from bottom to top at $T=0.5$)
$p=2$, $2.01$, $2.1$, $10$ and $3$.

\bigskip

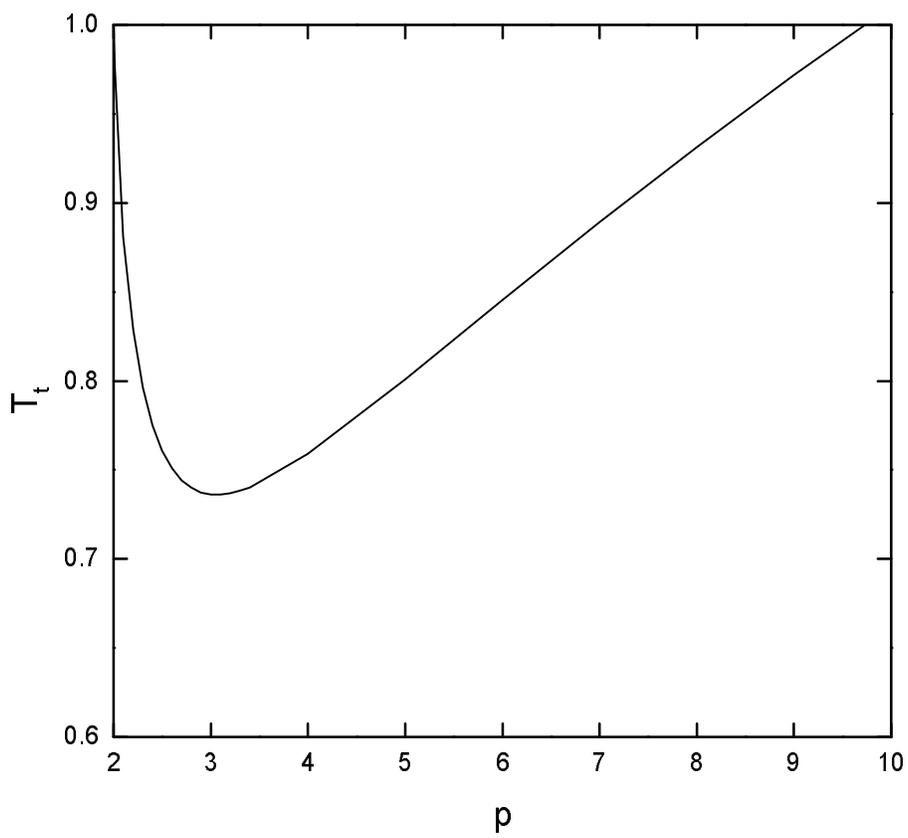
{\bf Fig. 8} Temperature at the tricritical point $T_t$
as a function of $p > 2$. Only integer values of $p$
have physical meaning.

\bigskip

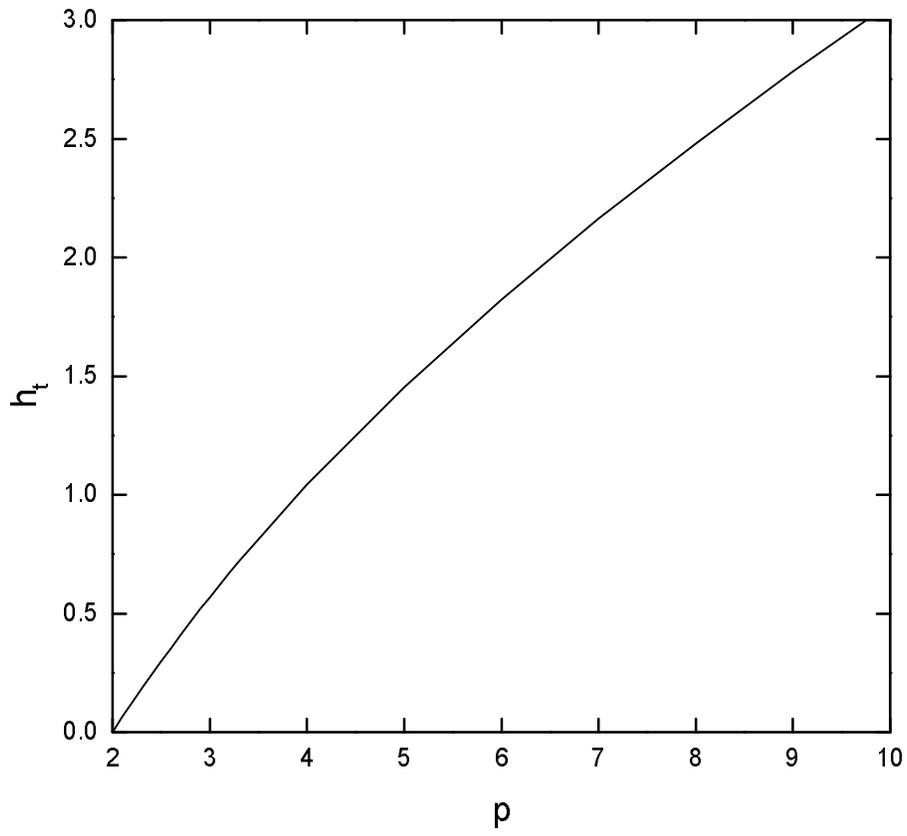
{\bf Fig. 9} Magnetic field at the tricritical point $h_t$
as a function of $p > 2$. Only integer values of $p$
have physical meaning.

\end{document}